\documentclass[superscriptaddress, twocolumn,
amssymb,amsmath,nobibnotes,aps,pra,showkeys,showpacs]{revtex4-1}
 
\usepackage{bm}
\usepackage{latexsym}
\usepackage{amsfonts,amssymb,amsmath} \usepackage{graphicx,epsfig}
\usepackage{psfrag} \usepackage{subfigure} \usepackage{color}
\usepackage{ulem} 
\newcommand{\be}{\begin{equation}} \newcommand{\ee}{\end{equation}}

\begin{document}

\title{Phase transition in a  Aubry-Andr\'e system with rapidly oscillating magnetic field}

\author{Tridev Mishra}\email{tridev.mishra@pilani.bits-pilani.ac.in}
\author{Rajath
  Shashidhara}\email{rajath.shashidhara@gmail.com}  \author{Tapomoy Guha
  Sarkar}\email{tapomoy1@gmail.com} \author{Jayendra
  N. Bandyopadhyay}\email{jnbandyo@gmail.com} 
\affiliation{Department
  of Physics, Birla Institute of Technology and Science, Pilani
    333031, India.}

\begin{abstract}
We investigate a variant of the Aubry-Andr\'e-Harper (AAH) model
corresponding to a bosonic optical lattice of ultra cold atoms  under an effective oscillatory
magnetic field. In the limit of high
frequency oscillation, the system maybe approximated by an effective
time independent Hamiltonian. We have studied
localization/delocalization transition exhibited by the effective
Hamiltonian. The effective Hamiltonian is found to retain the tight
binding tri-diagonal form in position space. In a striking contrast to
the usual AAH model, this non-dual system shows an energy dependent
mobility edge - a feature which is usually reminiscent of Hamiltonians
with beyond the nearest neighbour hoppings in real space.  Finally, we
discuss  possibilities of experimentally realizing this system in optical lattices.

\end{abstract}
\maketitle

\section{Introduction} 
Ever since its proposal by Anderson \cite{Anderson}, localization, and
transitions between localized and extended states, have been studied in a variety of systems \cite{Evers}.
Extensive analysis has been undertaken to understand various aspects of
metal-insulator transitions, localization as well as existence of
mobility edges in quasi-periodic or disordered 1D lattices using
scaling and renormalization techniques \cite{GrempelPrange, ostlundPandit, Kohmoto, Hiramoto89prb,
  Thouless,DassarmaXie88,90,Hashimoto92}. A system which has served as a rich
  prototype for such studies is the Hamiltonian, originally due to Harper \cite{Harper},
  and investigated for phase transitions by Aubry and Andr\'e \cite{AA}.

An important feature of the Harper Hamiltonian is the existence of a
metal-insulator transition \cite{AA} reminiscent of Anderson
transition. However, a notable difference is the absence of an energy
dependent mobility edge separating the localized and extended states,
which is a distinguishing feature of the Anderson transition in
3D. The Aubry-Andr\'e-Harper (AAH) Hamiltonian exhibits a sharp,
duality driven, transition at a unique critical value of the lattice
modulation strength for all energies \cite{Svetlana,
  belissardSimon,bsimon,AvronSimon}.  An ensuing trend in recent works
has been  to develop variations on the model which manifest a
Anderson-transition like mobility edge \cite{biddle,
  ganeshanDassarma}.

  The quest is legitimized further by the substantial progress made by
  the cold atom community in reproducing complex condensed matter
  phenomena including Anderson localization \cite{Billy, Roati}.

    The experimental investigation of localization in 1D systems,
    especially of the quasiperiodic/incommensurate crystalline
    variety has witnessed a sustained interest ever since such
    lattices could be realized using ultra cold atoms in a bichromatic
    optical potential, or photonic quasicrystals
    \cite{Drese,Roati,Lahini,Roux,Roth,Elye,Raizen,ModungoNJP}. These
    studies have ranged from direct experimental demonstration
    \cite{Damski,Roati,Guarrera,Ritt,Fallani,Lahini} to numerical
    calculations \cite{Roux,Roth,Elye} with accompanying proposals for
    observing the appearance of localized phases and the
    Metal to Insulator or superfluid to Mott insulator transition (in the
    presence of interactions). Here,
    the control on the degree of commensurability has helped to
    identify the point of transition which, in the AAH model is the
    self duality induced critical point \cite{ModungoNJP}. The
    Hofstadter variant and the AAH model in a 2D optical lattice have
    also been successfully realized \cite{ketterle, bloch}, in the
    context of simulating homogeneous magnetic fields in optical
    lattices.  The effects of periodic driving on localization
    phenomena in 1D disordered systems as a possible means of
    weakening localization and arriving at extended or non local
    states has shown encouraging results
    \cite{Yamada,DFMartinez}. Similar pursuits in AAH systems with a
    view to analyzing diffusive transport behaviour and wavepacket
    dynamics in the presence of driving, have been promising in terms
    of appearance of delocalized states \cite{Lima,Hufnagel,Kovolsky}.

The technique of 'shaking' of ultracold atoms in optical lattices has
risen to prominence as a flexible means of generating new effective
Hamiltonians which may replicate the effects of disorder, curvature,
stresses and strains, and several other phenomena as synthetic gauge
fields both Abelian or non-Abelian \cite{Drese,Arimondo,P.Hauke,tridev}.
 Some recent studies in driven cold atom setups have looked at induced
 resonant couplings between localized states thereby making them
 extended \cite{Morales} or at localization through incommensurate
 periodic kicks to an optical lattice \cite{QinYinShen}.  In these
 models, the phase transition instead of being driven by disorder, is
 a consequence of deliberate incommensurate periodicity. Demonstration
 of this behaviour has also been sought in a phase space analysis of
 the transition \cite{Hanggi, Wobst}.

  However, conspicuous by their absence have been works which look at
  the AAH model with a rapidly oscillating magnetic field, using the
  extensive tunability of cold atom setups.  The existing study of AAH
  systems assumes the magnetic field to be static in time. If the
  magnetic field is periodic, then one may find a perturbative
  solution in the limit of high frequency driving.  We address this
  neglected aspect by employing a formalism based on Floquet analysis
  to obtain an approximate effective time independent Hamiltonian for
  the system \cite{Grozdanov, Rahav, Maricq, Avan, Dalibard,
    Polkovnikov}.  A pertinent enquiry about the effective system
  would be to look for a metal to insulator phase transition with an
  energy dependent mobility edge.

 In this work, we consider a high frequency, sinusoidal effective
 magnetic field which couples minimally to the AAH Hamiltonian.  The
 effective Hamiltonian is obtained for this system and its
 localization characteristics are compared with the usual self-dual
 AAH model in real and Fourier space. An energy dependent mobility
 edge has already been studied in the context of an AAH Hamiltonian
 with an exponentially decaying strength of hopping parameters (beyond
 nearest neighbor coupling) \cite{biddle}.  We explore the possibility
 of a similar mobility edge in our physically motivated effective
 Hamiltonian with only nearest neighbor hopping. The non self-dual
 nature of our model is analysed and some general features are
 investigated.   In the section on Discussions, we attempt to
   reconcile our findings for the specific case with generic features
   of such non-dual models thereby putting the results in
   perspective. Finally, we discuss some possible experimental
   techniques that could be adapted to realize a version of the model
   presented here. Here, we highlight the difficulties involved in
   doing so and compare our model to some other driven cold atom AAH
   models in the literature.  
\section{Formalism}
Recent successes in synthesizing tunable, possibly
time-dependent, artificial gauge fields for systems of ultra cold
neutral atoms in optical lattices \cite{Struck,YJLin} has opened a
gateway to the strong field regime required for Hofstadter like systems
\cite{Hofstadter}. The system to be studied here may also be realized as an
incommensurate superposition of two 1D optical lattices
\cite{ModungoNJP}, with the laser beams for one of them undergoing a
time-dependent frequency modulation. This shall be discussed in detail later, under Experimental Aspects.

We consider a tight binding  Hamiltonian with nearest neighbor coupling that can
be expressed as a time dependent Aubry-Andr\'e-Harper Hamiltonian of the form
$H(t) = H_0 + V(t)$, where
 \begin{equation}
\begin{split}
\label{AAH}
 H_0 = &\displaystyle\sum_{n}\vert n \rangle \langle n+1 \vert + \vert n \rangle \langle n-1 \vert \\
V(t) = &V_0\sum_{n} 
 \cos \bigl[2\pi\alpha_0 n\cos(\omega t)  +\theta \bigr]\, \vert n\rangle \langle n \vert. 
 \end{split}
\end{equation}
The summation here runs over all lattice sites. The time-dependent
parameter $\alpha(t) = \alpha_0 \cos(\omega t ) $ denotes the flux
quanta per unit cell. An irrational value of $\alpha_0$ shall render
the on-site potential to be quasi-periodic. The harmonic time
dependence of $\alpha(t)$ owes its origin to a time dependent magnetic
field $ {\bf B} = B_0 \cos (\omega t) {\bf\hat z}$. The other
parameter $\theta$ is an arbitrary phase.  The $\lvert n\rangle$'s are
the Wannier states pinned to the lattice sites which are used as the
basis for representing the Hamiltonian and $V_0$ denotes the strength
of the on-site potential. The time dependence in the argument of the cosine modulation of the on-site potential is  different from usual time dependent AAH models where it is in the overall magnitude of the on-site potential. The periodic time-dependent operator $V(t)$
can be expanded in a Fourier series as
\begin{equation}
 {V}(t)= {\widehat V}_0 + {\displaystyle\sum_{1\leq j<\infty}\widehat{V}_{j}e^{ij\omega t}} + {\displaystyle\sum_{1\leq j<\infty}\widehat{V}_{-j}e^{-ij\omega t}}.
\label{eq:fourier}
\end{equation}
In order to obtain the effective time independent Hamiltonian one
writes the time evolution operator as 
\begin{equation}
U(t_{i},t_{f}) =
e^{-i\hat{F}(t_{f})}e^{-i{H}_{\rm eff}(t_{f}-t_{i})}e^{i\hat{F}(t_{i})},
\end{equation}
where, one introduces a time dependent Hermitian operator
$\hat{F}$. The idea is to push all the time dependence to the initial
and final ``kick'' terms and render the main time evolution to be
dictated by a time independent Hamiltonian. The systematic formalism
yields in the limit of large $\omega$ the following perturbative
expansion  for the effective time independent Hamiltonian given by 
\cite{Dalibard}
\begin{equation}
\begin{split}
\label{effective}
 {H}_{\rm eff} &= {H}_0 + \widehat{V}_0+ \frac{1}{\omega}{\displaystyle\sum_{j=1}^{\infty}\frac{1}{j}[\widehat{V}_j,\widehat{V}_{-j}]} \\
& + \frac{1}{2\omega^2}
 {\displaystyle\sum_{j=1}^{\infty}\frac{1}{j^2}\biggl(\biggl[[\widehat{V}_j,H_0],\widehat{V}_{-j}\biggr]+h.c. \biggr)} + \mathcal{O}(\omega^{-3}),
\end{split}
\end{equation}
 where, $\omega^{-1}$ is the  small perturbation parameter,  and the series is truncated at $\mathcal{O}( \omega^{-2})$.
In order to find the effective approximate Hamiltonian representing
our system in the large $\omega$ limit, one needs to compute the
Fourier coefficients in Eq.\eqref{eq:fourier}.  This is done by using
the following commonly valid expansions \cite{abramo}
 \begin{equation}
\begin{split}
 \cos(r\cos x) = \mathcal{J}_0(r)  + 2{\displaystyle\sum_{p=1}^{\infty}(-1)^p\mathcal{J}_{_{2p}}(r) \cos(2px)} \\
 \sin(r\cos x) = 2{\displaystyle\sum_{p=1}^{\infty}(-1)^{p-1}\mathcal{J}_{_{2p-1}}(r) \cos[(2p-1)x]}, \end{split} \end{equation}
where, $\mathcal{J}_{_n}(r)$ 's are Bessel functions of order $n$. The Fourier coefficients of ${V}(t)$ may be obtained by inverting  Eq.\eqref{eq:fourier} using these expansions. We obtain  
\begin{align}
\label{values}
 \widehat{V}_{j} = &(-1)^{\frac{j}{2}}V_0\cos\theta{\displaystyle\sum_{n}  \mathcal{J}_{_{j}}(2\pi\alpha_0n)}\,\vert n\rangle \langle n \vert;  ~~ j=\pm 2,4,6...\nonumber\\
 \widehat{V}_{j} = &(-1)^{\frac{j+1}{2}}V_0 \sin\theta{\displaystyle\sum_{n}\mathcal{J}_{_{j}}(2\pi\alpha_0n)}\,\vert n\rangle \langle n \vert;  ~~ j=\pm 1,3,5...\nonumber\\
{\widehat V}_{0} = &V_0 \cos\theta{\displaystyle\sum_{n}  \mathcal{J}_{_0}(2\pi\alpha_{_0}n)}\,\vert n\rangle \langle n \vert.
\end{align}
We find that  $[\widehat{V}_{j},\widehat{V}_{-j}] = 0$  owing to the symmetric nature of the Fourier coefficients (for
real V). Therefore the $\mathcal{O}(\omega^{-1})$ correction to the
effective Hamiltonian vanishes and the first non-trivial correction is at  $\mathcal{O}(\omega^{-2})$. The
$\mathcal{O}(\omega^{-2})$ term of the effective Hamiltonian would require the 
 commutator bracket $ \biggl[[\widehat{V}_j,H_0],\widehat{V}_{-j}\biggr]$, which  on evaluation yields

\begin{widetext}
\begin{equation}
\label{offsite}
\biggl[[\widehat{V}_j,H_0],\widehat{V}_{-j}\biggr]=\begin{cases}\begin{split}
 \displaystyle\sum_{n} - V_0^2\cos^2\theta  \biggl[\biggl(\mathcal{J}_{_{j}}[2\pi\alpha_0 &(n+1)] - 
\mathcal{J}_{_{j}}(2\pi\alpha_0n)\biggr)^2 \vert n\rangle \langle n+1\vert \\
&+ \biggl(\mathcal{J}_{_{j}}[2\pi\alpha_0(n-1)] -\mathcal{J}_{_{j}}(2\pi\alpha_0n)\biggr)^2 \vert n\rangle \langle n-1\vert\biggr]
 \end{split} & \textrm{ if $j=\pm 2,4,6...$}\\
\begin{split}
\displaystyle\sum_{n} - V_0^2\sin^2\theta  \biggl[\biggl(\mathcal{J}_{_{j}}[2\pi\alpha_0 &(n+1)] - 
\mathcal{J}_{_{j}}(2\pi\alpha_0n)\biggr)^2 \vert n\rangle \langle n+1\vert \\
&+\biggl(\mathcal{J}_{_{j}}[2\pi\alpha_0(n-1)] -\mathcal{J}_{_{j}}(2\pi\alpha_0n)\biggr)^2 \vert n\rangle \langle n-1\vert \biggr]
 \end{split} & \textrm{ if $j=\pm 1,3,5...$}
\end{cases}
\end{equation}
\end{widetext}
Using the above expression in Eq. \eqref{effective} we obtain the
effective Hamiltonian, ${H}_{\rm eff}$, for our system. We find that
up to $\mathcal{O}(\omega^{-2})$, the effective Hamiltonian yields a
nearest neighbor tight binding model with a zeroth order Bessel
function modulating the site energies, and higher order Bessel
functions make their appearance in the hopping terms. There have been 
works which have looked at inhomogenities in the hopping of the AAH model, arising not from driving
but from the choice of next nearest neigbour hoppings in the corresponding 2D quantum hall model\cite{Claro,Hatsugai,thouless,Han}.
However, these models consider situations where the off-diagonal modulations are quasiperiodic through incommensurate modifications
of cosine kind of terms. In our case above, the incommensurability is embedded in higher order Bessel functions, thereby variations
in hopping strength are far more erratic and with signatures bordering on those of disorder. This is expected to have ramifications
for the localization/extended behaviour of the eigenstates. This has been discussed in the next section  and illustrated 
through localization phase plots.
\begin{figure}
\includegraphics[height=3.2cm,width= 7cm]{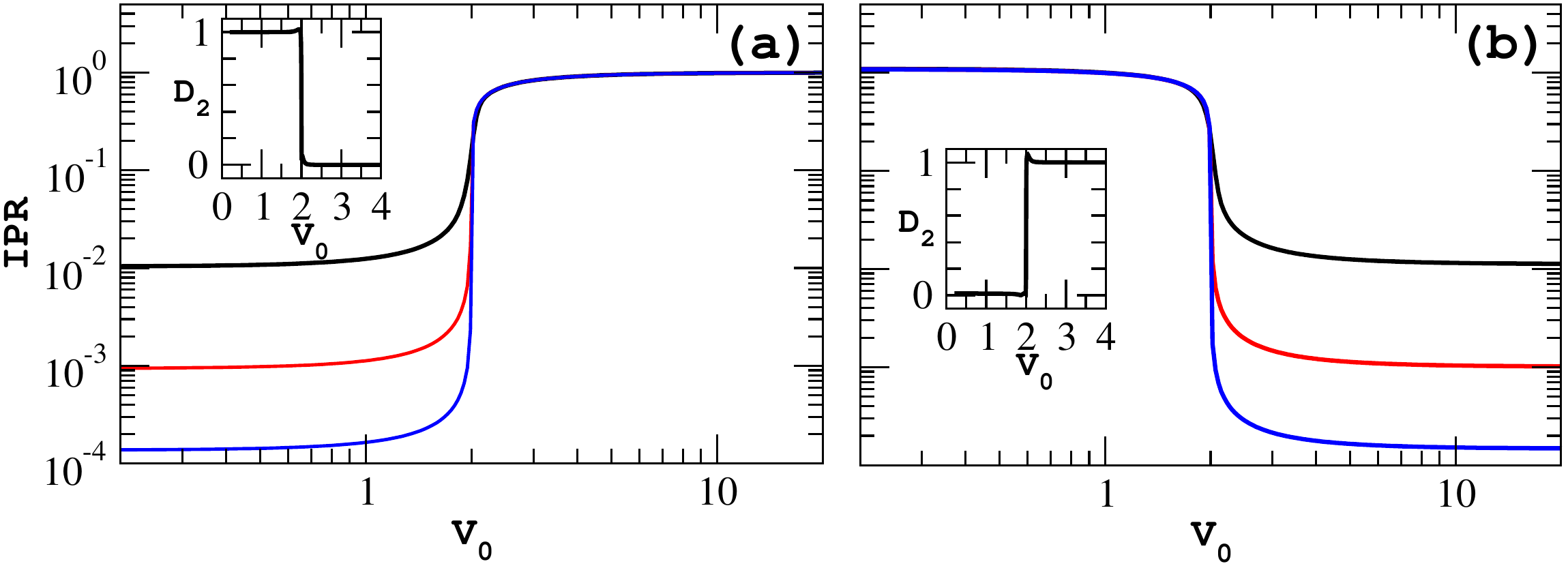}\\
\includegraphics[height=3.2cm,width= 7cm]{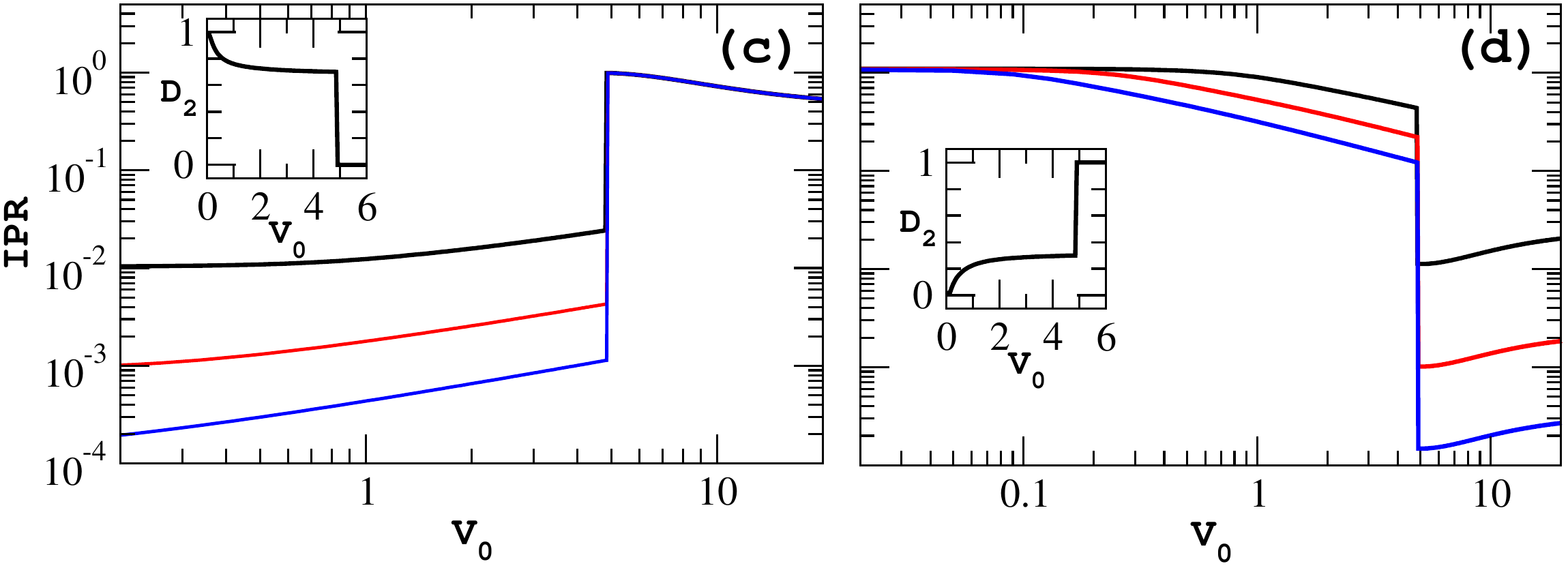}
\caption{(Upper Panel)The metal to insulator transition of
  the AAH Hamiltonian for the  system's ground state. Plot (a) shows the IPR versus $V_0$  in real
  space for  $L =144,1597 ~\textrm{and}~
  10946$ (top to bottom). The inset shows the variation of  $D_2$ with $V_0$ which also exhibits a transition. Plot (b) exhibits the mirror behaviour in the dual
  space. (Lower Panel) The transition seen in  the IPR versus $V_{0}$ curve  for the $H_{\rm eff}$. Plots (c) and (d) are the real and dual space plots 
   for the  ground state with phase $\theta = 0$.}
\label{fig:AAFAAipr}
\end{figure}

\section{Results} 
The simple AAH model has a well studied transition from
localized (insulating) to delocalized (metallic) phase which occurs at
a critical value $V_0 =2$. To
quantify the localization property we use the inverse participation
ratio (IPR) defined as $\textrm{IPR} =
{\displaystyle\sum_{n=1}^L|a_n|^4}/\left({\displaystyle\sum_{n=1}^L|a_n|^2}\right)^2$
where $a_n$'s are the expansion coefficients of the energy eigenstates in a local discrete site basis and $L$ the number of lattice
sites \cite{ThoulessPhysRep,BellDean}. The IPR takes a value in the range $1$ to $1/L$ with $1$
indicating a perfectly localized state and $1/L$ for completely
extended states.  Figure \ref{fig:AAFAAipr}(a) shows the transition
in the real space IPR for the ground state of the AAH system with choice of
irrational $\alpha_0$ as  inverse of the golden mean $(\sqrt{5}-1)/2$ and $L =144,1597 ~\textrm{and}~
  10946$.  The
inset in this plot indicates variation in the magnitude of the quantity $|D_2|$ with $V_0$, where  $\textrm{IPR} \propto L^{-D_2}$. 
For a given $V_0$, $D_2$ is obtained by fitting a linear regression line between $\log \textrm{IPR}$ and $\log L$ and thereby obtaining the slope. The regression fitting is  done using  several values of $L$, taken to be large Fibonacci numbers. In all the plots discussed here the transitions depicted are 
for some finite choice of system size and hence not exactly 'step' changes but ramp up or down over some finite range of $V_0$ values. Further, the 
references to such transitions as abrupt or occuring at a critical value have to be interpreted within such numerical constraints.
$D_2$ values shows an abrupt transition from $1$ to $0$ at the critical value
irrespective of lattice size.  This establishes the transition to be a
integral feature of the model even in the thermodynamic limit of
infinite lattice size and the critical  point is protected in
this limit.
In order to switch to  states in the Fourier domain, i.e, $\vert m\rangle$ 's from the position space kets $\vert n\rangle$
we use the transformation
\begin{equation}
 \vert m\rangle = \frac{1}{\sqrt{L}}\displaystyle\sum_{n}\textrm{exp}(-i 2\pi m\alpha_0 n)\vert n\rangle.
\end{equation}
 This enables one to write the AAH Hamiltonian in Fourier
space and compute the IPR in this space. Figure \ref{fig:AAFAAipr}(b) 
shows the transition in Fourier space for a set of parameters
identical to those in plot (a).  Here, again the characteristic
transition occurs at the critical point $V_0 =2$ and  the
curves in plot (a)  are a mirror reflection of the curves in plot (b) about $V_0 =2$. This is due to the exactly self dual nature of the AAH Hamiltonian. Thus,  an
extended regime in real space implies a localized one in Fourier space
and vice versa. The inset for $D_2$  in Fourier space accordingly
mirrors its real space counterpart.

In the case of our effective model, the IPR for  the ground state exhibits a similar trend, as shown in
Fig.\ref{fig:AAFAAipr}(c).  The transition in this case for this
state occurs at a new critical value $V_0\approx4.6$, all parameters
being kept the same as in former plots. Another distinguishing feature
of this transition for the driven case is the manner in which the IPR
values approach the critical value and depart from it, relative to the
behaviour in the standard AAH model. In the extended regime, instead
of the perfectly flat value of ${\rm IPR }\approx 0$, a slow positive
gradient is observed indicating weak localization that progressively
gets stronger until a sharp surge occurs at the critical point. Even
beyond the transition there is a fall in the IPR due to the still
imperfect nature of the localization. This unique behavior can be
partially attributed to the non self-dual nature of the effective
Hamiltonian which shall be discussed later.  $D_2$, in the inset,
continues to retain its scale invariant attributes and manifests the
imprint of the transition. The difference from the simple AAH model,
lies in the fall in its value from $1$ to a lower plateau before the
transition. An indication of the existence of a parameter regime where
the state is neither purely localized nor extended but a sort of
composite i.e. critical and possibly multifractal. This is due to the unusual behavior of the wavefunction in
the case of $\alpha_0$ being a Liouville irrational number whereby,
for some lattice modulations, no finite localization length may be
found over which the state could be said to appreciably decay
\cite{bsimon,AvronSimon}.  Plot (d) which shows the Fourier space IPR
for our model exhibits reciprocal behavior of the kind seen in the
simple AAH model but with the major difference that plots (c) and (d)
are not mirror reflected about the same critical value. This deviation
is expected on grounds of the non self dual nature of our system's
Hamiltonian. In Fourier space, $D_2$ analogously is not an exact
mirror image of its real space version but all other qualitative
characteristics remain the same.
There are features in the driven system's Fourier space IPR which stand in contrast
from the AAH model, as seen in plots (b) and (d) of
Fig.\ref{fig:AAFAAipr}. Most notably, the driven model shows a
discrimination between the different lengths as the curves in plot (d)
transition from localized to extended regimes at different rates. This
is not the case in the ordinary AAH model, where all lengths transition
together, as seen in plot (b). This difference is an indicator of non-nearest neighbor couplings in  our dual space effective Hamiltonian
and the accompanying anomalous behavior of the wavefunctions in a certain parameter range. 
\begin{figure*}[t]
\begin{tabular}{lcr}
\includegraphics[height=5.5cm,width=6.2cm]{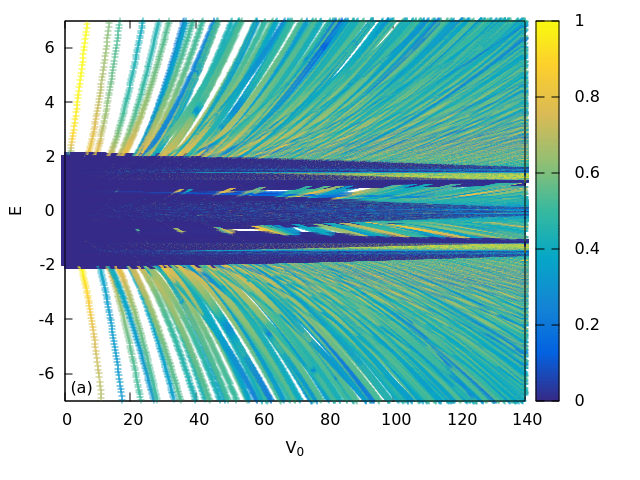}&
\includegraphics[height=5.5cm,width= 6.2cm]{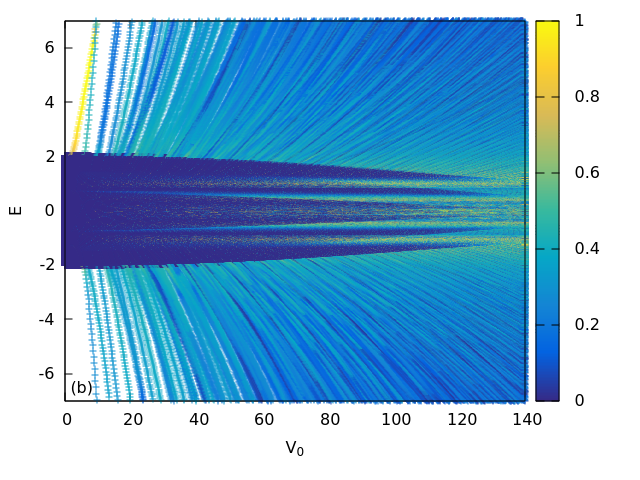}&
\includegraphics[height=5.5cm,width=6.2cm]{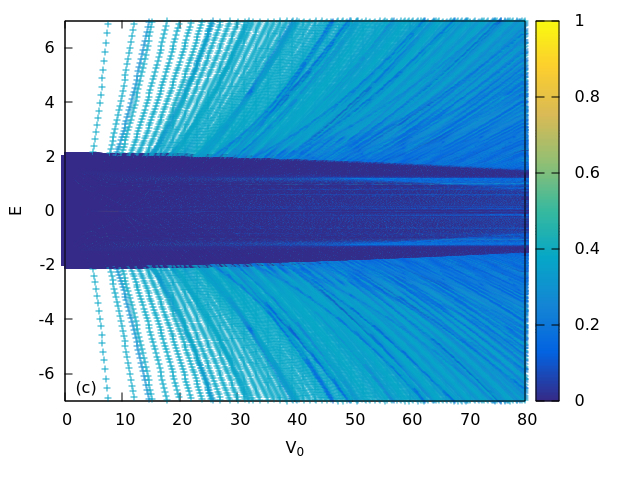}
\end{tabular}
\caption{ (Color online) The localization phase diagram with  IPR in the (E,$V_0$)
  phase plane, for lattice size $L=4181$ three values of the variable $\theta$ (a)
  $\theta = 0$ , (b) $ \theta = \pi/4$, (c) $\theta = \pi/2$. }
\label{fig:mobedge}
\end{figure*}

In order to understand how the properties of the transition are
  related to the normal modes of the effective driven system, we look
  at the localization phase diagram, i.e.  the variation of IPR with
  $V_0$ and the low energy region of the spectrum at each $V_0$
  . Figure \ref{fig:mobedge} shows the IPR in the $V_0 - E$ plane.  We
  consider three such plots for values of the phase, $\theta = 0$ ,
  $\theta = \pi/4$ and $\theta = \pi/2$ in Eq.\eqref{AAH} and lattice size $L= 4181$. The nature
  of the variation of IPR reveals a sharp energy dependent mobility
  edge for our model. The portion of the energy spectrum for which the
  IPR variation has been illustrated is chosen to clearly indicate the
  appearance of localized states . The choice of values for the phase
  is intended to isolate and compare the relative effects of the
  modified onsite term and the site dependent hopping terms. In all
  three plots the sector corresponding to low $V_0$ values and near to
  $E=0$ shows a dense region of extended states which, (see eq.\eqref{AAH}), reflects the bare hopping
  structure.

The features in the these phase diagrams owe their origin to the relative dominance of  different terms 
the driven effective Hamiltonian for a given $\theta$. 
The anisotropy of the zero-order Bessel modulated on-site
  energy adds impurity/disorder like effects on top of the inherent
  quasiperiodicity of the model.  The fall off of this on-site energy
  with lattice sites diminishes the actual system size to a reduced
  one. The modified hopping strengths are also site dependent and vary
  in an oscillatory manner, with a damping with increasing 
  site index. When the onsite potential damps out, the hopping terms from $H_0$  survive.
   This is expected to contribute to an increase in the IPR
  as the behaviour tends to one of a lattice with disorder. These
  factors collectively influence the spectral spread and density
  alongwith the loclalized/delocalized behaviour of the various
  eigenstates.

 Plot \ref{fig:mobedge} (a), for $\theta = 0$, depicts the appearance of
  quasi-localized states (yellow fringes) at the boundaries of the
  extended (deep blue) region. As $V_0$ increases, the dense region
  of extended states near $E=0$ begins to manifest traces of localization
  in IPR values, introducing the mobility edges. This can be noted
  from the bifurcations of the phase boundary that begin to show up in
   with increasing $V_0$ with localized eigenstates piercing into portions of the spectrum which
  at lower $V_0$ were dominated by extended states. For  higher
  energy  the IPR values vary primarily between critical and
  extended behaviours. This is notably absent in the plots for $ \theta = \pi/4$ and $\theta = \pi/2$ where critical behaviour is hardly observed and
  that too in a very narrow region around the phase boundary.
 The wider gapping in the eigenvalues as compared to the other two cases can be accorded to the
  overall stronger influence of the on-site term as compared to the
  hoppings.

 Plot (b), for $\theta = \pi/4$, includes the effects of all the terms of the effective Hamiltonian.
 The appearance of localized states around $E = 0$ takes place as before.
 However, there is a notable lack of appearance of well localized states in the higher energy regions as compared to plot (a).  
This indicates a closer competetion between the on-site and hopping terms of the driven model. The significant localization effects, hence mobility edges,  appear 
distinctly  in a band around $ E = 0$  and that too at higher $V_0$. This may be accounted for by the fact that for $\theta = \pi/4$ both
the sine and cosine  modulations are present in the modified hopping (see eq.\eqref{offsite}), unlike the previous $\theta = 0$ case. Apart from contributing to the predominance of extended
states in most of the spectrum, this more influential hopping part also makes the spectrum relatively less gapped.

Plot (c), for $\theta = \pi/2$ has been specifically shown to illustrate how the hopping terms in the driven AAH model behave in 
the absence of any on-site term. For this  choice of $\theta$  the ${\widehat V}_{0}$ in eq.\eqref{values} goes to zero.
As expected in the presence of just the hopping,  the IPR values show an extended behaviour everywhere in the phase plot. However, one
can still note a phase boundary differentiating the region of the purely extended states from somewhat less extended ones. The appearance
and nature of the bifurcations in the boundary of this dense part of the  phase plot with changing $V_0$,  indicates the qualitative effects of the 
inhomogeneities in the hopping terms. 
Looking at  plots (a) (b) and (c) it is clear that the role of the modulated onsite has the effect of enhancing the localization of states as well as 
creating a sharper mobility edge.
 
This indicates that the driven model shows a strong sensitivity to the phase $\theta$  in terms of the localizaton behaviour and the appearance of mobility edges.
The number of these edges, as can be seen, is more for the $\theta= \pi/4$ case and is almost  absent in the phase plot for $\theta=\pi/2$.  A recent work \cite{Liu}, looks at a topological classification of AAH
models with  cosine modulated hoppings which differ by a phase factor from the onsite modulation. This helps to realize topologically distinct families
of AAH Hamiltonians with the possibility of topological phase transitions between the different classes via a modification of the lattice modulations.
Similar behaviour would be interesting to study in our driven context. 

\section{ Discussion } 
In order to qualitatively analyze some of our results we  consider a simplified model comprising of a trivial constant hopping term and an on-site potential $T$ which is aperiodic or quasi-periodic. The Schr\"odinger equation is given by  
\begin{equation}
\label{eq:AA}
 a_{n+1} + a_{n-1} + \Lambda  T(\alpha_0 n + \phi)a_n= Ea_n
\end{equation}
where,  $\Lambda$ is the strength of
on-site energy and $E$ are  the energy eigenvalues. One can go to the dual
space for the above system by defining an expansion, as follows
\begin{equation}
\label{trans1}
 a_n = \frac{ e^{ikn}}{\sqrt L}\displaystyle\sum_m \tilde{a}_m
 e^{i\,m(\alpha_0 n + \phi)}
\end{equation}
where, the $\tilde{a}_m$'s are the dual space amplitudes, and $k$ is a wave vector
from the Bloch wave expansion ansatz.  This  allows 
 $T$  to be expressed as
\begin{equation}
\label{trans2}
  T(\alpha_0 n+ \phi) = \frac{1}{\sqrt L}\displaystyle\sum_{\acute{m}}\mathcal{T}_{\acute{m}} e^{i\,\acute{m}(\alpha_0 n + \phi)}.
\end{equation}

Equations \eqref{trans1} and \eqref{trans2}  yield an on-site term in the dual space from the the hopping terms of Eq.\eqref{eq:AA} as  
\begin{equation}
\label{eq:Fspacesite}
a_{n-1} + a_{n+1} = \frac{e^{ikn}}{\sqrt L}\displaystyle\sum_m \tilde{a}_m e^{i\,m(\alpha_0 n + \phi)} \cos(\alpha_0 m + k)
\end{equation}
with a cosine modulation of the on-site energy  (as seen in
the Aubry and Andr\'e model). Interestingly, the real space on-site energy term transforms as
\begin{equation}
\Lambda  T(\alpha_0 n + \phi)a_n= \frac{\Lambda e^{ikn}}{L}\displaystyle\sum_{\acute{m}}\displaystyle\sum_m\mathcal{T}_{\acute{m}}\tilde{a}_m e^{i\,(m + \acute{m})(\alpha_0 n+\phi)}.   
\end{equation}
The RHS can be slightly rearranged to give 
\begin{equation}
\label{farTB}
\begin{split}
\frac{\Lambda e^{ikn}}{L}&\displaystyle\sum_{\acute{m}\neq\pm1}\displaystyle\sum_m\mathcal{T}_{\acute{m}}\tilde{a}_{m + \acute{m}} e^{i\,m(\alpha_0 n+\phi)} +\\ 
& \frac{\Lambda e^{ikn}}{L}\displaystyle\sum_m (\mathcal{T}_1\tilde{a}_{m-1} + \mathcal{T}_{-1}\tilde{a}_{m+1})e^{i\,m(\alpha_0 n + \phi)} 
\end{split}
\end{equation}
In the above form,  the second term clearly indicates the apparent nearest
neighbor hopping terms in the dual space whose strength is modulated
by the Fourier components of $T$. It is the first term in the above
expression which explicitly breaks the exact duality. The form of
$\mathcal{T}_{\acute{m}}$ determines the extent to which different
$m$ values in the dual space are coupled.  It is well known
that for decaying oscillatory functions like Sinc and Bessel function
of the zeroth order $\mathcal{T}_{\acute{m}}$ is a rectangular
function,   with possibly a $\acute{m}$ dependent modulation,  symmetric
about the origin. Thus,  in our case, we expect a truncation effect in
dual space which restricts the range of couplings. This deviation from
exact duality is expected to have some impact on the
probability of an ``analytic accident'' along the lines of \cite{AA}.
The appearance of localized states (real eigenfunctions) happens when
there are superpositions of counter-propagating plane waves with
wave vectors of near-commensurate magnitude. This would mean,  in our
model, some harmonics from the expansion of $T$ shall scatter the wave with
wave vector $k$ by an amount commensurate with $2n\pi$. This has to be
considered together with the fact that for a rational approximation of
$\alpha_0$ as a ratio of two large successive Fibonacci numbers, the
true momentum(Fourier) space eigenvalues $\kappa$ are related to $m$
as $ \kappa = mF_{i-1}\textrm{mod}F_i$, where $F_{i-1}$ and $F_i$ are
successive Fibonacci numbers \cite{Kohmoto,Hanggi}. Thus, what appear
to be close neighbors in $m$ could possibly be well separated in the
actual wave vector space. Further, the range of $m$ values that shall
remain coupled in the dual space will be dictated by the extent of
$T$ in the real lattice for example the first zero in the Bessel
function. The set of $m$'s which conspires with a given $k$ value to
yield a localized state shall be dictated by $V_0$ and $E(k)$. This
explains the energy dependent mobility edge in Fig.\ref{fig:mobedge}. 

In the dual space, where $k$ acts as a phase (see Eq.\eqref{eq:Fspacesite}), a
state localized at few $m$ values could be  shifted by large amounts
for a small change in $k$. This allows for the interpretation that a
small change in $\phi$ could in effect cause a state localized around
some lattice site to localize about a far off site.
 In terms of symmetry, the absence of 
translational invariance in Euclidean space of quasi-periodic
structures with two incommensurate periodicities can be restored in an
extended space using the $\phi$ dimension \cite{Sokoloff, Janner}. This
effect of $\phi$ on localization properties leads to the differences between the
three plots in Fig.\ref{fig:mobedge}.

\section{Experimental Aspects}

The experimental realization of our system may be achieved in several
different ways, with ease and feasibility of implementation being the
guiding criteria in the choice of method. We will explore two options
here, from recent literature,  which are promising. One
way is to begin with a 2D optical lattice and then proceed in the
manner described in some recent realizations of the Harper-Hofstadter
Hamiltonian \cite{ketterle,bloch}. The notion of simulating a synthetic
magnetic field by means of generating effective flux per plaquette of
the lattice is a generic feature. However, the true appeal of these
methods compared to others for generating artificial magnetic fields
for ultracold neutral alkali metal atoms in optical lattices, is the
absence of coupling between different hyperfine states of the
atoms. It is possible therefore to proceed with a single internal
state and far detuned lasers to achieve homogeneous magnetic fields by
a laser assisted hopping process. A pair of far detuned Raman lasers
is employed, while tunneling in a particular direction is obstructed
by means of a gradient/ ramp in the site energies using gravity or
magnetic fields, to restore resonant tunneling between sites. The AAH
Hamiltonian is obtained in a time independent effective way by
averaging over the high frequency terms and the hopping energy is
modified by a complex position dependent phase.

We suggest using Raman lasers of frequencies close to those of the
optical lattice lasers, as prescibed by the authors, and to use the
tunability of the flux per plaquette $\mathbf{\alpha_0}$ offered by
such choice, to set it to an irrational value by adjusting the angle
between the Raman beams. Introducing the time dependence is admittedly
tricky. This is due to the fact that the static Harper Hamiltonian in
the above technique is itself achieved by time averaging and we need
it to have a further residual time dependence.  For this one would
have to vary $\mathbf{\alpha_0}$ periodically by say, modifying the
angle between the Raman lasers periodically with time together with
simultaneous time modulations of the detunings and the gradients in a
fashion that the overall effect is of a periodic change that is of a
rate slower than the oscillations to be averaged over while resonant
tunneling occurs, but fast enough to remain detuned from the energy
gap between the ground and excited bloch bands of the trapped atoms in
the lattice potential. This yeilds a time scale which survives the
first averaging and gives one a time dependent AAH Hamiltonian
effectively being driven by an oscillating magnetic field. There are
some obstacles to be overcome here such as arranging the time
dependent detunings and the angular variation of the Raman lasers so
as to vary $\mathbf{\alpha_0}$ sinusoidally as a function of time
effectively, since there is a good chance of getting high frequency
noisy components that have to be averaged over. Another issue is that 
 the scheme does not realize the simple Landau gauge
for a magnetic field. We use this gauge in our analysis but the
results, essentially the nature and existence of the Metal-insualtor
transition, are independent of any such choice through the gauge
freedom embodied in the choice of $\mathbf{\theta}$ in the AAH
Hamiltonian\cite{AA}. The analysis then would be modified only upto a
gauge transformation. It would indeed be interesting if the method
could be modified to include the Landau gauge.

On account of the multiple time dependent modulations in the
realization just discussed, heating and spontaneous photonic emission
processes are a legitimate source of concern. We would like to outline
another approach using a quasiperiodic 1D optical lattice which may
have better characteristics as regards dissipative processes. Here we
suggest using the bichromatic 1D optical lattice realization of the
AAH model as described in \cite{ModungoNJP} and suitably modifying it
to implement our model. Essentially, a bichromatic optical lattice
setup is one with a pair of superposed standing waves wherein one
provides the tight binding structure to the Hamiltonian and the other,
a weak secondary perturbing potential which, through adjustable
non-commensurability of its wavelength with that of the first, offers
a quasiperiodic/pseudorandom potential for the ultracold gas of atoms
even to the extent of mimicking quasidisorder in the lattice
\cite{Roati}. The two standing waves have wavelengths in the ratio of
two consecutive Fibonacci numbers. This helps to realize a workable
notion of incommensurability in a finite lattice system by tending the
value of the $\mathbf{\alpha_0}$ to near the inverse of the golden
mean. As suggested in \cite{ModungoNJP} this is the key requirement
for the observation of a transition from extended to localized states
i.e. to keep a large number of lattice sites in a single period of the
on-site potential for finite systems.

The next step is to systematically introduce the driving. This is done
by introducing a time dependence in the ratio of the wavelengths of
the two standing wave lattices. More precisely, to do this we suggest
generating the two standing waves using beam splitting and retro
reflection by mirrors. If now the reflecting mirror corresponding to
the primary tight binding lattice is shaken according to a protocol
which mimics a sinusoidal drive say, by mounting it on a piezoelectric
motor, it should be in principle possible to generate a sinusoidal
time dependence in the irrational flux term. It would be preferable to
use actuators that move the mirrors so as to produce acceleration
effects on the lattice such that one may achieve time dependent
Doppler shifts in the frequency and hence wavelengths of the
stationary waves (where one averages over the fast oscillations in the
amplitudes to get the hopping terms) which could be controlled in a
sinusoidal fashion. This may be technically demanding under the
present capabilities of shaking in optical lattice systems but is
definitely worth exploring as a powerful instrument for studying
effective Hamiltonians in a new time dependent regime.

In the two approaches highlighted above, the respective works
\cite{ketterle} and \cite{ModungoNJP} provide a clear map between the
parameters of the simulated model and the experimental parameters such
as laser intensities, recoil energies of the trapped neutral atoms and
the energy gap between the ground state and lowest excited bloch band
in the lattice. This mapping translates readily to the formalism of
calculating the effective Hamiltonian. For instance in the case of the
bichromatic construction the mapping of the continuous optical
potential to a tight binding picture has been carried out in
\cite{ModungoNJP} using a set of local Wannier basis states. As per
this construction our time dependent model, see eq.\eqref{AAH} would
have $\mathbf{V_0}$ to be a ratio of product of the height of the weak
perturbing lattice with the time dependent ratio of the wavelengths of
the two standing waves and an integral term, to the hopping element of
the primary optical lattice. The term $\mathbf{\alpha_0}$ is just the
ratio of the wavelengths of the two standing waves, made time
dependent by shaking, which are two consecutive Fibonacci
numbers. From the expressions in eqs. \eqref{values} and
\eqref{offsite} it can be readily seen how the experimental parameters
enter into the modified on-site and nearest neigbour hopping energy
terms of the high frequency effective static Hamiltonian for the
driven system.

Thus the relation between parameters of the setup and the derived
model Hamiltonian can be traced in a straightforward manner. It is
true that this manner of constructing the system will make the
strength of the on-site modulation (or its ratio with the hopping
energy) also a sinusoidal function of time but this is not expected to
alter the system's features studied here in any significant way. We
propose studying this more general form of time dependence, with the
strength of the on-site to off site energy also taken to be a function
of time alongside the periodicity in $\mathbf{\alpha_0}$, as a future
line of work.

Briefly, we would like to survey how our work contrasts with other
early and recent work, which has emphasized periodic driving through
additional potentials \cite{Morales} and shaking \cite{Drese} in AAH
systems. In \cite{Drese}, shaking introduces a time dependent phase in
the cosine term of the on-site energy. This phase is seperate from the
incommensurate position dependence. The effect is a renormalization of
the hopping energy so as to make it a function of the driving
amplitude.  This enables one to tune across the metal-isulator
transition by varying the amplitude of the driving. Whereas in our
system the driving is provided through an oscillatory effective
magnetic field which manifests itself through the periodicity in
$\mathbf{\alpha_0}$, hence present in the incommensurate position
dependent term.

This is again different from \cite{Morales} which employs a driving
that is a weak space quasiperiodic and time periodic perturbation onto
the AAH system modeled as a quasiperiodic optical lattice. Here, the
driving is a weak perturbation to the original AAH Hamiltonian. In our
case, however, the manner in which the AAH model is driven is
non-perturbative by its very nature.  An oscillating magnetic field,
even of small amplitude, is in no way a weak perturbation and cannot
be treated as such, it has to be looked upon in the Floquet picture of
periodic time dependent Hamiltonians. The high frequency nature of the
driving permits a Floquet theoretic treatment of a slightly analytical
variety through the high frequency expansion available for the Floquet
Hamiltonian. Only here, in the parameter $\mathbf{1/\omega}$, is one
allowed to use a perturbative treatment. This is formally different
from \cite{Morales} in that our modifications significantly alter the
AAH model for which there is limited analytical footing in the high
frequency regime. It would do well to regard the newly obtained static
effective Hamiltonian as an independent system in its own right, with
features that are not to be expected in the undriven or weakly driven
AAH models. This in fact merits looking into, as it would not be
unjustified to anticipate exotic modifications to the traditional AAH
Metal Insulator transition under these circumstances.

Also worth noting are the differences between the model in
\cite{Morales} and our system from a reciprocal space point of
view. While our model is also non self dual it has an exact 1D
structure with couplings that are beyond the nearest neighbour. In
\cite{Morales}, the dual Hamiltonian is not exactly 1D and the
extended states appear due to resonant couplings of localized states
that are driving induced. This differs considerably from the mechanism
(discussed in the previous section)that causes
localization/delocalization behaviour in our driven system. In fact,
the non self duality of our model sets it apart even from undriven
variations on the AAH model (with mobility edges) which are self dual
by construction \cite{biddle,ganeshanDassarma}, irrespective of the
real space couplings being nearest neighbour or beyond it.  

\section{Conclusion}
In this work we have studied the Aubry- Andr\'e-Harper problem with an
oscillatory magnetic field in the promising cold atom- optical lattice scenario. The
problem is significantly simplified by going into an effective
Hamiltonian which approximately represents the system in the limit
of  high frequency magnetic field. We find
that this effective Hamiltonian is non-self dual, and though it exhibits a
metal-insulator transition, it differs from the classic Aubry-Andr\'e
model in the emergence of an energy dependent mobility edge. The
nearest-neighbor coupling form of the effective Hamiltonian yields
this feature which is commonly observed in  disordered 3D systems or
Aubry-Andr\'e like models with hoppings extending beyond the  nearest
neighbor.\\

\section{Acknowledgements}
T.M. would like to acknowledge Takashi Oka and Alessio Celi  for valuable discussions on various theoretical and experimental aspects
of the problem.

\end{document}